# MATISSE: a novel tool to access, visualize and analyze data from planetary exploration missions

Angelo Zinzi*[a,b], Maria Teresa Capria[c], Ernesto Palomba[c], Paolo Giommi[a], Lucio Angelo Antonelli[a,b]

a) ASI Science Data Center, c/o ASI, Via del Politecnico snc, 00133, Rome, Italy

b) INAF-OAR, Via Frascati n. 33, 00078, Monte Porzio Catone (RM), Italy

c) INAF-IAPS, Via del Fosso del Cavaliere n. 100, 00133, Rome, Italy

* Corresponding author: Angelo Zinzi. Address: ASI Science Data Center, c/o ASI, Via del Politecnico snc, 00133, Rome, Italy. Email: angelo.zinzi@asdc.asi.it. Phone: +39 068567884

**Abstract**

The increasing number and complexity of planetary exploration space missions require new tools to access, visualize and analyse data to improve their scientific return.

ASI Science Data Center (ASDC) addresses this request with the web-tool MATISSE (Multi-purpose Advanced Tool for the Instruments of the Solar System Exploration), allowing the visualization of single observation or real-time computed high-order products, directly projected on the three-dimensional model of the selected target body.

Using MATISSE it will be no longer needed to download huge quantity of data or to write down a specific code for every instrument analysed, greatly encouraging studies based on joint analysis of different datasets.

In addition the extremely high-resolution output, to be used offline with a Python-based free software, together with the files to be read with specific GIS software, makes it a valuable tool to further process the data at the best spatial accuracy available.

MATISSE modular structure permits addition of new missions or tasks and, thanks to dedicated future developments, it would be possible to make it compliant to the Planetary Virtual Observatory standards currently under definition. In this context the recent development of an interface to the NASA ODE REST API by which it is possible to access to public repositories is set.

**Keywords:** Web data visualization, Planetary Sciences, 67P/Churyumov-Gerasimenko, Solar System Exploration, Joint Analysis, Virtual Observatory

## 1. Introduction

In the last decade the number of space missions dedicated to planetary exploration sensibly raised, covering all type of objects in the Solar System (i.e., planets, satellites, dwarf planets, asteroids and comets) and acquiring very different types of observations.

Without information about, for example, observation conditions or positioning, performing scientific analysis of a planetary body would likely need the download of a huge quantity of data to be subsequently finely selected and processed by means of customized software.

Another issue regards the correct use of data, not straightforward even for those present and documented in public repositories. This makes it not trivial for a research team specialized in a given field to correctly manage data acquired by instruments not belonging to that field (e.g., observing a different spectral range).

And, in a sector where the cross matching of different types of observations is of great value, the limited capability of merging data acquired over the same target by different instruments constitutes a major obstacle.

These issues could be solved by using a database with metadata information (e.g., geographic position or observing geometry) together with a system that automatically reads and processes observations acquired by different instruments and missions.

The astronomical community started to remove this kind of obstacles by developing the Virtual Observatory (VO – e.g., Quinn et al., 2004; Hanish, 2014), assuring access to the data without requiring files on the local machine or writing different libraries for every dataset and research study.

Given the intrinsic different and varied nature of data of planetological interest, efforts to adapt this standard to planetary sciences only recently began (Erard et al., 2014). Hence planetary science is still mainly based on single-instrument analysis, or to "ad-hoc" collaborations between a limited number of instruments (e.g., Keihm et al., 2012).

In this context the ASI Science Data Center (ASDC) launched its plan related to the Solar System Exploration in the late 2012 (Zinzi et al., 2014), aiming at giving the community a tool to be used with a wide range of missions and instruments over different targets, with high-order products as output.

This project can be now considered at a noteworthy milestone with the version 1.1 of the multi-platform, web-based tool, named MATISSE (Multi-purpose Advanced Tool for the Instruments for the Solar System Exploration), available online at http://tools.asdc.asi.it/matisse.jsp.

MATISSE currently ingests both public and proprietary data from 4 missions (ESA Rosetta, NASA Dawn, Chinese Chang'e-1 and Chang'e-2), 4 targets (4 Vesta, 21 Lutetia, 67P Churyumov-Gerasimenko, the Moon) and 6 instruments (GIADA, OSIRIS, VIRTIS-M, all onboard Rosetta, VIR onboard Dawn, elemental abundance maps from Gamma Ray Spectrometer, Digital Elevation Models by Laser Altimeter and Digital Ortophoto by CCD Camera from Chang'e-1 and Chang'e-2). However, it is worthy to note that its modular structure easily allows to embrace a wider range of instruments and targets from past, present and future missions.

In addition the NASA ODE REST API (Bennett et al., 2014) is now used to add the possibility of searching through remote archives: at the present time only one instrument is available with this option (MDIS-NAC onboard NASA's MESSENGER), waiting users feedbacks to expand this functionality.

This paper, devoted to the description of the MATISSE tool and its capabilities is structured as follows: Section 2 illustrates the workflow behind MATISSE, in Section 3 the database used in MATISSE will be presented, in Section 4 some use cases will be shown and, finally Section 5 is for conclusions and future developments of the tool.

## 2. Workflow description

MATISSE is designed using a wide spectrum of computer languages:

- a Java query interface collects input parameters
- these are passed to the appropriate bash shell script
- the actual workflow is launched by this latter script, so that the final products are displayed on the web and downloadable by the user.

Other languages used are IDL (by means of GDL – GnuDataLanguage - the free compiler for IDL), Python and C.

The first turning point is that regarding the choice between single- or multi-observation mode (hereafter referred to as SOM and MOM, respectively).

The differences between these two modes mainly reside in the beginning and in the end of the workflow. On the contrary the central part of the process, dedicated to data reading and interpolation with the (unevenly distributed) shape model grid, is only repeated for every observation required by the user (one for SOM, various for MOM).

### 2.1 Web interface and input parameters

The homepage of MATISSE allows a first search of observations in the database relying on target, mission, instruments, and (not mandatory) geographic (latitude and longitude respect to the selected target) and geometric parameters (emergence, incidence and phase angles of observations). The query starts when the "Search" button is clicked (Fig. 1).

[FIGURE 1]

The result of this query will be displayed in tabular form and the user can select the desired observations by looking at the values of the different columns displayed regarding each of them (e.g., Start Time, Longitude Min or Latitude Max - Fig. 2). For SOM it is needed to click on a row, select colour table and levels (and wavelength, if present) and finally click the "Submit" button. For the MOM, after each observation selection the user must click on the "Next" button and only when all the desired observations have been loaded the "Submit" button has to be clicked. In this mode also the operation to be applied, between average, ratio, difference (these two require exactly 2 observations) and RGB (that requires exactly 3 observations), must be selected. At this point the workflow proceeds to its bash shell phase.

[FIGURE 2]

As described in Table 1 the SOM case accepts 25 parameters, while the MOM case 26 for the "Submit" button and other 6 for the "Next" button.

[TABLE 1]

### 2.2 Bash shell starting scripts

The back-end of MATISSE is made of two bash shell scripts, namely matisse.sh and matisse_multi.sh, for SOM and MOM respectively.

The first one uses the input variables from the web interface to select the required observation, read it using the dedicated software (generally provided by the instrument team) and finally interpolate the original data to the shape model grid.

The second script will create several scripts depending on the number of observations selected: matisse_manager.sh and one matisse_<timeID>.sh (where <timeID> identifies a unique time code) for each observation required by the user.

These scripts are identical to matisse.sh and will be used to read and interpolate original data to the final grid, whereas matisse_manager.sh will merge the interpolated data to compute the result required by the user, applying the operation chosen.

### 2.3 Data reading and interpolation

The most part of public data from planetary exploration missions are in PDS (Planetary Data System – JPL, 2009) format and stored together with the software to read them (generally written in IDL); data still not public are deposited in different formats, but the code to read them is often distributed in the (restricted access) archive (usually in IDL language).

Where present, in MATISSE we use these libraries to read the data archived in the ASDC database: for example some VIRTIS-M data are read by means of virtispds software (Erard et al, 2013).

#### 2.3.1 Instrument acquiring surface related data

A bash shell script, named <instrument>.sh (where <instrument> is the name of the required instrument), creates a Linux executable named run_gdl_<instrument> that finally opens the GDL environment launching the script <instrument>.pro.

At the end of this process the original data is stored in binary format and passed as input to idw.sh, a bash shell script that, after creating idw.pro with the correct parameters, launches it in the GDL environment to interpolate the original data to the shape model grid.

The idw.pro script selects the data on the basis of the input parameters provided by the user and interpolates them thanks to some Python codes, making use of the griddata function from scipy.interpolate package (Jones et al., 2001): at the present time a linear interpolation method is applied, after the boundaries of the area to be analysed have been identified with a nearest neighbour interpolation (exploiting the fact that points outside this area have NaN values).

Therefore the original observation, with values acquired on a non-regular latitude/longitude grid defined by the instrument Field Of View characteristics, is transformed to the projected grid of the shape model or digital elevation model using as coordinates those from the pixel center of the observation.

This interpolation is made possible by means of observing geometries (i.e., values of latitude, longitude, emergence angle and similar parameters for every observed pixel) present for every file: even if these are generally given with the observation file (in form of a separated file, inside the PDS label or as a layer of the image cube) in few cases they need to be calculated by us before insertion in the database. This is the case for the OSIRIS images of Lutetia, that are distributed on the PSA archive without observing geometries and for which we computed them using the NASA's NAIF Spice Toolkit in order to associate to every observation present in the database a file with the corresponding geometric information.

The entire process is depicted in Fig. 3.

**[FIGURE 3]**

The same process is applied by the scripts matisse.<timeID>.sh to each observation selected by the user in the MOM case, with the addition of another IDL script, launched by the executable run_idw_multi that, applying the requested operation (average, ratio, difference or RGB), merges the output of each idw.pro, giving the final result (Fig. 4)

The final output of this process is a binary file containing only the resulting data in floating point format, covering the whole geographic range of the target body.

Other binary output files are the ones with data values regarding only the selected area, the latitudes and longitudes belonging to this area and the colour-converted (RGB) data.

ASCII files are those describing the minimum and maximum values of the interpolated data, the RGB colorbar used in the output page and RGB specification of the IDL colour table used (all these files are not distributed to the user, but only used to generate the output page).

**[FIGURE 4]**

### 2.3.2 Instrument acquiring non surface related data

This workflow is different for the data acquired by GIADA (Colangeli et al., 2007; Della Corte et al., 2014a) because, differently from the other instruments ingested by MATISSE, it collects data belonging to coma and not to surface characteristics: therefore the visualization will be made for points over the orbit of the spacecraft.

The difference in the workflow is the IDL script needed to create the output (not idw.pro, but giada_x3d.pro) and in the computation on the fly of observation geometries with the executable spice_giada, written in C using the NASA NAIF Spice Toolkit (Fig. 3).

It is however worthy to note that up to now only GIADA simulated observations (Della Corte et al., 2014b) are present in the MATISSE database and, therefore, also the geometries are related to the old and obsolete shape model of the 67P used for their computation.

### 2.4 Web-based data visualization

The files generated by the process described above are then read by the Java code used to build the output page (Fig. 5), together with some parameters, such as the target body to be shown.

For every target body available in MATISSE two shape models with different resolutions are generally present: the lower resolution one is for the web-based visualization, the higher resolution is for the offline desktop visualization.

Shape models are generated from those released with the NASA Spice kernels or from Digital Elevation Models (triangulated with a Delaunay method). The file used in the output web-page to assign a colour code to every vertex of the target contains an array of floats with three columns (for red, green and blue) and the same number of rows of the points present in the corresponding shape model.

**[FIGURE 5]**

This univocal correspondence is needed to display the 3D shape model coloured according to the values of the data requested by the user. The three-dimensional, interactive visualization is based upon the x3dom JavaScript (X3DOM, 2016), freely available online, that allows the user to rotate, zoom and click on the target body, so that also information about the value at a precise point can be retrieved (this last option is only available for the MOM).

Other two visualizations are present in the output page, namely two-dimensional projected maps of the data, one "local" (i.e., zoomed over the geographical extension requested by the user) and one showing the entire target projected.

These maps are generated by a Python script that uses SciPy and Matplotlib (Hunter, 2007) modules, creating as outputs the two PNG images shown in the web-page and two downloadable PostScript files, with higher resolution and suitable to be used for science publications.

The colour table is the same chosen by the user between the 40 available in IDL and is exactly replicated in Python, so that the same color-coding is present in both 3D and 2D visualizations.

### 2.5 Downloadable data

Apart from the data shown, in the output web-page also downloadable files are available, allowing the user accessing higher-resolution visualizations: all the downloadable files are packed together in two TGZ archives reachable by clickable links at the bottom of the page.

In the first archive the files are:

- a GeoTIFF file created for GIS applications such as ArcGIS or QGIS generated taking into account the spatial reference (SR) system of the target body as listed on the "Spatial Reference" website (Spatial Reference, 2016 - not present for 67P and Lutetia[1])
- a couple of files (extension IMG/hdr) to be read with ITT ENVI software: in this case the three bands of the files are latitude, longitude and value
- a FITS file
- a "readme" file, explaining how to open the downloaded files

In the other compressed archive is present a vtp file, created by a Python script with VTK package (VTK, 2016), to be open with the open source, freely available online, Paraview software (Ahrens et al., 2004 – www.paraview.org). The same "readme" file contained in the other TGZ is present here.

The vtp file is of great importance since it uses the high-resolution shape model present in the MATISSE bundle and therefore far better visualization and analysis could be performed respect to the online version.

In addition Paraview is Python-based and can be used to perform further analysis on the data, such as comparison between different products and statistical analysis.

For non-surface observations a different output is provided in the (single) TGZ archive: in this case an IDL function, capable of reading the binary file attached, is present.

## 3. Database description

---

[1] The spatial reference systems here used are: 1) IAU2000:30100 for the Moon, 2) IAU2000:19900 for Mercury, 3) IAU2000:2000004 for Vesta

MATISSE uses a MySQL database to allow geographic/geometric queries: this database is made of several tables, describing the fundamental characteristics of targets, missions and instruments, the privileges of all the MATISSE users, and containing some crucial metadata about every single observation present in the archive.

The first dropdown menu the user has to choose from is that for the target selection, reading the "Target" table; subsequently missions available for this target are shown in the following dropdown menu (with values read from the "Missions" table). These information are used to read in the "ObservationType" table to search for instruments belonging to the selected target and mission together to those stored in "UserGroupObservation" to specify what observations the user can access (i.e., if the user belongs to some private group or only the public one): in this way only the instruments that can be accessed by the user are displayed in the checkbox menu created on the fly.

Clicking on the "Search" button a query inside the "ObservationMetaData" table starts, using the geographic/geometric parameters passed by the user (or the default ones if left blank) and, when the query results are displayed and one observation selected, the possibility to select the desired wavelength/channel (only for multi-channel instruments such as VIRTIS-M and VIR) is made possible by reading the "Spectrum Range" table.

### 3.1 Metadata table

The metadata table, named "ObservationMetaData", has been adapted from the parameters used in the EPNcore data model and EPN-TAP protocol developed in the frame of the Planetary VO activities in Europlanet programs (Erard et al., 2014): all the 20 parameters are described in Table 2, but the most important ones are latitude/longitude ranges and phase, emission and incidence angle ranges, together with the acquisition time information (respectively parameters 9-10, 7-8, 17-18, 15-16, 13-14 in Table 2). When a parameter included in the table is not available for a dataset, the value assigned is "NULL".

These fields are computed on the basis of observation geometries associated to every observation (given by data distributor or computed by us, as in the case of OSIRIS for 21 Lutetia described in Sec. 2.3.1).

Using tables similar to those proposed by the developers of the Planetary VO it would be easier in the future to use also their protocols (such as EPN-TAP), in order to expand the usability of the tools for data inside Planetary VO.

In order to remove the possibility of mismatch between the different table some parameters of ObservationMetaData are linked to the other ones inside the MATISSE database in the following way (e.g., the observationTypeId column is the Id of the ObservationType table)

[TABLE 2]

The synchronization of the internal archive with external repositories, such as ESA PSA and NASA PDS, is performed periodically with an IDL script that uses rsync to find new (or new versions of) observations not present in the database and IDL/C scripts to read/compute observational parameters to be included in the table.

It is important to note that, as data from missions of interest will be added to public repositories, MATISSE will be ready to include them in its database, so that new analysis could be performed.

[TABLE 3]

### 3.2 User privileges management

As already briefly stated above, MATISSE is able to manage both public and proprietary data, allowing only users with special privileges to access the latter. This goal is achieved by defining groups to which observations belong: public observations will always belong to the "anonymous" group, whereas private data will be part of a specific group (i.e., ROSETTA:VIRTIS or DAWN:VIRVMT).

By joining the information contained in the "UserGroupObservation" table (where every single observation present in the database is linked to a specific group using its observation_id) together with the users present in every group it is therefore possible to fulfil the desired requirement.

The addition of a user to a specific group is made after contacting the Principal Investigator (PI) of the instrument involved: therefore every user, just after registration to the MATISSE tool, will belong only to the "anonymous" group and only after confirmation by the PI will be added to the requested groups.

## 4. Use cases

After the complete description of the algorithms and procedures that allow MATISSE to perform its tasks, the present section aims at showing some of the capabilities of the tool, by means of use cases.

In particular here we will show the visualization of a single observation (Sec. 4.1 – OSIRIS/NAC for asteroid Lutetia), a mosaic made by OSIRIS/WAC images of Lutetia (Sec. 4.2), the difference between two simulated temperatures on the 67P Churyumov-Gerasimenko comet (Sec. 4.3), the RGB false colour visualization computed with the different wavelengths of VIRTIS-M VIS instrument (Sec. 4.4), the visualization of the GIADA data, not acquired on the surface (Sec. 4.5) and the high-resolution imaging output from the Moon (Sec. 4.6).

Apart from the lunar one, these cases have been selected on the basis of their data policy, using the public ones or those referring to simulations no more updated.

These data are very limited in number inside the MATISSE database that, on the other hand, contains thousands of VIRTIS-M observations of the comet 67P Churyumov-Gerasimenko, still not released to the public and, therefore, only accessible by the VIRTIS team and not usable in this work (Tab. 3).

[TABLE 4]

### 4.1 Single image visualization

The first and most immediate use of MATISSE is that related to the visualization of a single observation, such as the visible wavelength ones acquired by OSIRIS/NAC (Keller et al., 2007) onboard Rosetta spacecraft during its Lutetia's fly-by.

Even if this mode does not actually create a high-order product as output, it would be precious to visualize data directly on the shape model, especially when (as in the case of minor bodies) the surface is far from being considered spherical.

As shown in Tab. 4 it is possible to access this kind of data by selecting the appropriate Target/Mission/Instrument combination and then clicking on the "Search" button.

After selecting the (unique) desired observation, by clicking on the "Submit" button the SOM pipeline starts (see Sec. 2).

The output page (Fig. 6) will display on the left an interactive window where the selected image is overlaid to the 3D shape model of the target and on the right two static images, one zoomed on the geographic range of the image and the other showing the entire target in two dimensions.

Besides these graphic windows four clickable links are available, so that the user can separately download the two compressed archived files and the two high-resolution PostScript images related to the two dimensional maps shown in the page.

[FIGURE 6]

### 4.2 Average / Mosaic

The first MOM task applied to the data shown here is their averaging or, in other words, mosaicking, a task of great interest when a single observation is not able to cover a large portion of the area of interest.

In this case it would be important to include in the mosaic only data with a certain degree of similarity in what concerns observation conditions, otherwise results could be misleading. In fact at the present time MATISSE performs no photometric correction, basing it only on the preprocessing that scientific teams made on the data before the publication.

We illustrate this task by selecting the OSIRIS/WAC (Keller et al., 2007) images acquired over Lutetia with the F13 filter, with a selection of pixels with incidence angle less than 45° (Table 4).

After the available data list appears it is possible to choose the F13 observations by showing the filenames and then by clicking on each desired observation and on the "Next" button (see Sec. 2.1). After all the observations have been loaded to the system the MOM task can be selected ("Average") together with the colour table and the number of colour levels, finally clicking on the "Submit" button.

The output page is similar to that already shown in the previous section (Fig. 7): this time the data visualized on the shape model (and also shown in the 2D maps and available for the download) are not from a single image, but from all the observation selected and averaged together. In particular the average operation is applied to area of the "union" (i.e., not overlap/intersection) of the selected observations.

[FIGURE 7]

### 4.3 Difference

The possibility of computing the difference of two observables represents a diffuse way of analysis of planetary data.

For this example we will use theoretical surface temperature of the nucleus of 67P as obtained from a thermal model (Capria et al., 2014). In these simulations temperatures referring to the surface (first layer) and two layers at different depths under the surface (1 cm and 5 cm for second and third layer, respectively) are computed; in this case we will compute the difference between the first and the second layer ones.

The user starts selecting the comet 67P as target and SIMVIRTISD1V0 as observation. By performing a simple query, without filling any search field several observations will be found (Table 4).

At this point an observation can be selected and, after selecting the first layer temperature as measurement type the "Next" button must be clicked. The same has to be made for the second layer, after that the operation to be applied ("Difference") and, eventually, the colour table and the number of colour levels can be selected.

Clicking the "Submit" button the pipeline is launched and in the output page the shape model of the comet is shown, with the desired output overlaid (Fig. 8). As in the other cases already described the two static PNG images are displayed, together with the links to the PostScript files and to the compressed archives for download.

**[FIGURE 8]**

### 4.4 RGB false colour visualization

An interesting option to analyse three different observations is to load them as RGB channels, so that one is mapped as red, one as green and one as blue. In this way white colours are related to areas where the three observations have similar values, whereas areas where red/green/blue is predominant indicate a prevalence of the respective observation.

Respect to the difference described above, in this case the opportunity of comparing three observables rather than two is provided.

In our case we applied the RGB mode to the VIRTIS-M VIS (Coradini et al., 2007) radiance observation of Lutetia present in the MATISSE database, choosing the channels so that a "real colour" visualization will be available. To do this we selected the red channel around 700 nm, green at 546 nm and blue at 455 nm (Table 4).

Differently from the other MOM options (average, difference and ratio) for the RGB the colour table and colour levels selections have no effects, since the colouring is based on the RGB palette computed on the basis of the inputs and the number of levels is fixed to 256.

In this particular case the output page shows a "reddish" surface with some slight variations over the target body, thus indicating that the red channel is brighter than the other two (Fig. 9).

**[FIGURE 9]**

### 4.5 Non-surface data

The use case here shown is that regarding the visualization of data not coming from the surface, such as GIADA measuring the particle flux around comet 67P Churyumov-Gerasimenko (see Sec. 2.3.2).

After selecting this comet as target, Rosetta as mission and GIADA-Flux as instrument it is also possible to select acquisition time start and stop, time step (acting on the number of points shown), integration time and particle bin size (Table 4 - in this example we did not change the default values).

Differently from the other output pages already discussed, now only the interactive 3D window is present (together with the link to download data, as described in Sec. 2.5), showing the acquisition points all around the target body following the colour code chosen (Fig 10 left and center).

In addition orbital data are also projected on the surface of the target body using the normal to the surface from each point (Fig. 10 right). Even if the boresight direction, instead of the normal to the surface one, could be considered a better indicator for the computation of the surface projection, the data here used have not information about it, since they are only simulation computed long before Rosetta ended its hibernation phase.

When real data from GIADA will be added to the MATISSE archive it will be also possible to study the feasibility of producing movies from their time series, probably taking into account the work already done by the developers of 3Dview for the visualization of the coma (Gangloff, 2015)

[FIGURE 10]

### 4.6 High-resolution data from the Moon

Moon data currently ingested by MATISSE are derived products available on the website of the Chang'e missions (http://moon.bao.ac.cn/) and downloadable after registration and login.

In this section we will show how to access and visualize Digital Ortophoto Models (DOM), computed starting by data acquired by the CCD Camera onboard Chang'e-2 and with nominal resolution of 50 meters/pixel.

The web visualization is based on a 4 ppd (pixel per degree) global model of the Moon (Fig. 11), computed by NASA's Lunar Orbiter Laser Altimeter (LOLA – Smith et al., 2010), freely available on the PDS website at http://pds-geosciences.wustl.edu/lro/lro-l-lola-3-rdr-v1/lrolol_1xxx/data/lola_gdr/cylindrical/img/ (filename ldem_4.img).

The procedure to visualize lunar data is almost identical to that described in Sec. 4.1 for the OSIRIS/NAC observation of Lutetia (Table 4) and only the output page is different. In this case there will be only one static image, interpolated with the high resolution DTM used for offline purposes.

The NASA file used in this case changes with the selected area, since the 512 ppd map used is made of 32 sections (file names starting with ldem_512 and with img extension).

The global image is not present since the little geographic extension of areas of interest in this case does not allow the visualization on global projections. (Fig. 11).

[FIGURE 11]

The Paraview file, together with the ENVI and GeoTIFF ones, contained in the two downloadable compressed archives is the interpolation of the DOM data to the similar resolution global DEM provided by the PDS (Fig. 12).

[FIGURE 12]

# 5  Conclusions and future developments

A stable version of MATISSE (v 1.1) is now available online, aiming to be a starting point for joint data analysis in planetary exploration research.

This browser-based tool provides the Planetary Science community higher-order products, such as mosaics, ratios, differences and false colour (RGB) visualizations without the need of writing dedicated software nor downloading huge quantity of data.

Of great importance is the introduction of the possibility to project data directly on the three-dimensional model of the target, helping in visualizing where the observations have been acquired, in particular for the minor bodies with strongly irregular shapes (e.g., comet 67P Churyumov-Gerasimenko).

In addition, the extremely high-resolution final products downloadable for desktop software can be analysed in great detail, further extending the applications of the tool and allowing joint analysis of data acquired by different instruments or missions at the maximum resolution available.

All these features can be also easily applied to other datasets of planetological interest still not ingested by MATISSE, as the addition of both new Moon data from the Chang'e-2 mission and MESSENGER MDIS-NAC observation of Mercury in the last months demonstrate.

In the very next future we are planning to add public observations of Vesta acquired by the VIR instruments onboard NASA Dawn, as well as its observations of Ceres (even if with restricted access as no public data are already available).

Albeit the MATISSE version described in this paper can be certainly considered a stable and really usable one, the work to upgrade and develop its capabilities is far from ended. One of the major issues to be studied is that regarding the database to use, as MySQL is maybe not the best suited for GIS applications compared to, for example, PostgreSQL (coupled with PostGIS) or MongoDB.

It is also worthy to note that the core of the algorithm developed to create the high-resolution output has been already used for particular studies, where data currently not present in the MATISSE database have been projected on the 67P Churyumov-Gerasimenko shape model.

This is the case, for example, of the visualization of valuable surface characteristics of the 67P nucleus as inferred by Rosetta-VIRTIS spectrometer (Capaccioni et al., 2015). Or even, of the high resolution 3D images used to better understand the geology of Vesta (Palomba, et al., 2015; Longobardo et al., 2015).

For this reason it is already planned to extend this skill, distributing a multiplatform desktop application, allowing projections of data not included in the MATISSE database in formats often used in the planetological community (i.e., ENVI, GeoTIFF).

Even if the ability of browsing data through external repositories is already present in the present day online version of the tool, as the data from the MESSENGER mission demonstrate, this last application, together with the planned future use of EPN-TAP protocol, could be capital for the accomplishment of the goal related to a more strict interface with the Planetary-VO, for which MATISSE (and its desktop application under development) would represent an ideal 3D extension.

## Acknowledgments

We would like to thank Simonetta Dari, Gianni Di Eugenio and Fabrizio Fabri from Telespazio for their work in supporting MATISSE, in particular by developing the Java web front-end and the database realization.

**FIGURE CAPTIONS**

Fig. 1: The upper part of the MATISSE homepage, where input parameters have to be supplied by the user.
Fig. 2: The MATISSE homepage showing the query results.
Fig. 3: Workflow of the Single-Observation-Mode
Fig. 4: Workflow of the Multi-Observation-Mode (only the effect of "Submit" button is shown)
Fig. 5: The output page of MATISSE showing an OSIRIS/WAC observation of 21 Lutetia. The image on the left is the 3D interactive output, whereas on the right the 2D static projections (zoomed over the selected area on top; global at the bottom) are present. Above the "Back" button also the link to download high-resolution data and PostScript images is present.
The shape model used has been computed starting from that described by Sierks et al. (2011) and named LUTETIA_098K_CART.WRL
Figure 6: MATISSE output page for the SOM case. Images and links in the page are the same as described in Fig. 5. Also the shape model is that described in Fig. 5.
Figure 7: Result of the mosaicking. Images and links in the webpage have the same ordering as in Fig. 5
Figure 8: Output page of the difference operation. Images and links are as in Figs. 7 and 9.
Figure 9: Output page of the RGB visualization for VIRTIS-M VIS observations.
Figure 10: Visualizations for the GIADA instrument.
Figure 11: The output page for the DOM data. On the left the 3D visualization at 4ppd, on the right the 2D projected data (selected area is too small to be visible at this resolution).
Figure 12: The Paraview visualization of the file created with MATISSE, ranging 2°x2° in latitude/longitude

**Figure 1**
**Click here to download high resolution image**

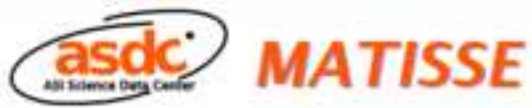
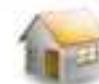

**Figure 2**
**Click here to download high resolution image**

**Figure 3**
**Click here to download high resolution image**

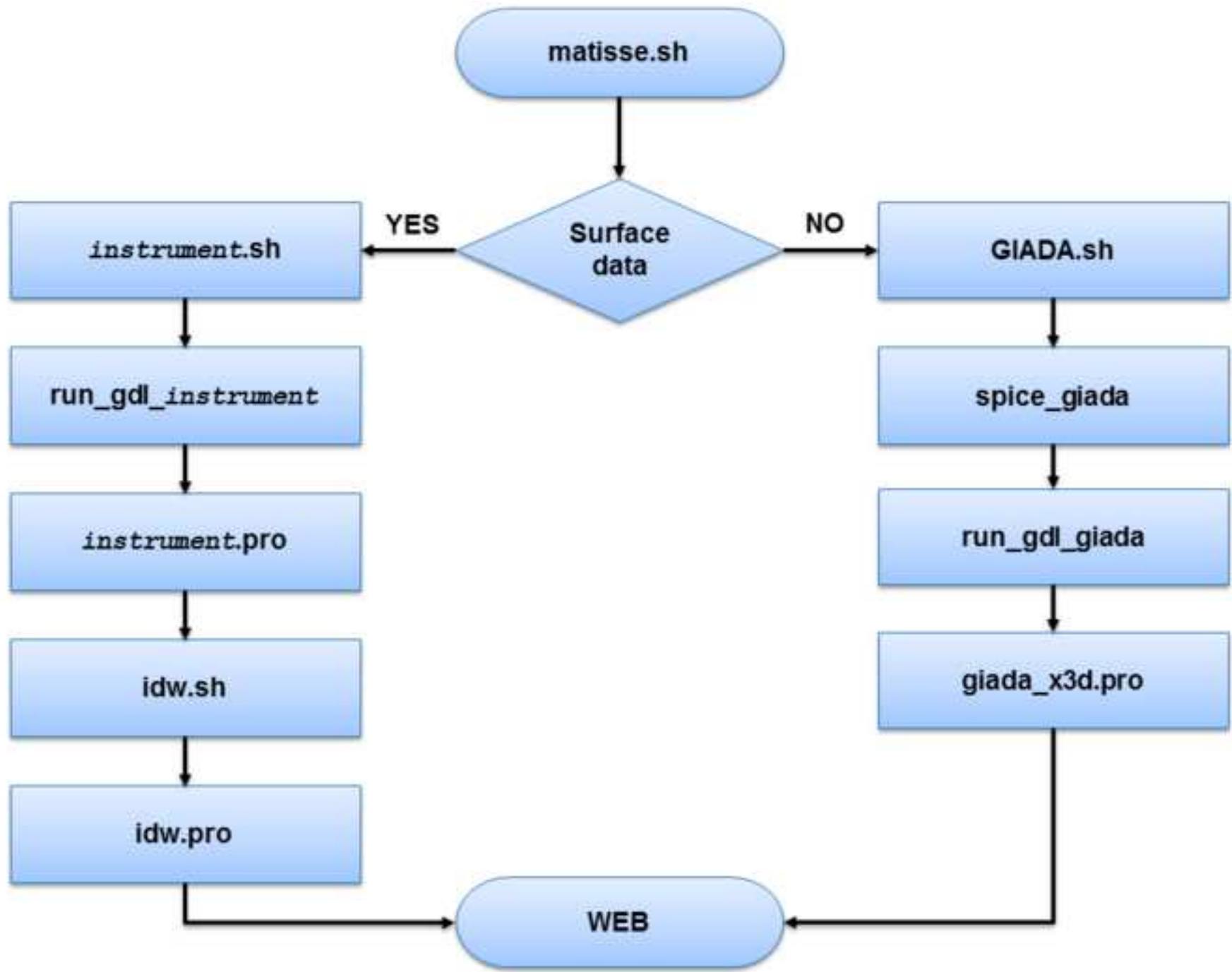

**Figure 4**
[Click here to download high resolution image](#)

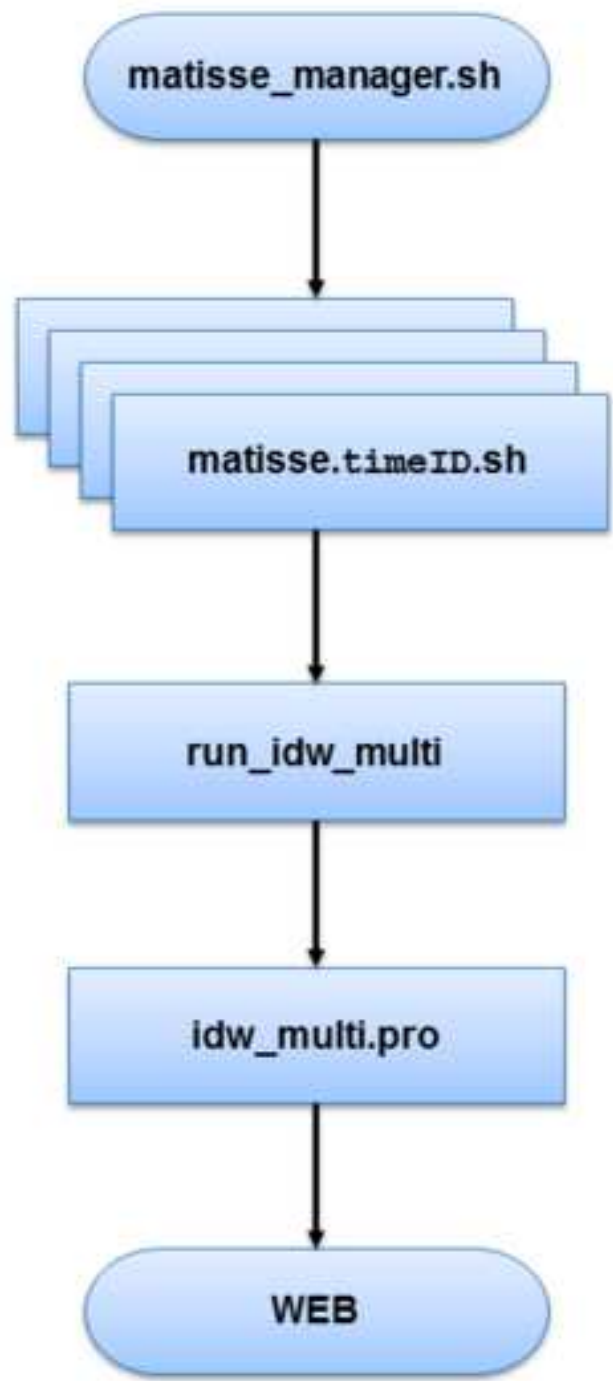

**Figure 5**
**Click here to download high resolution image**

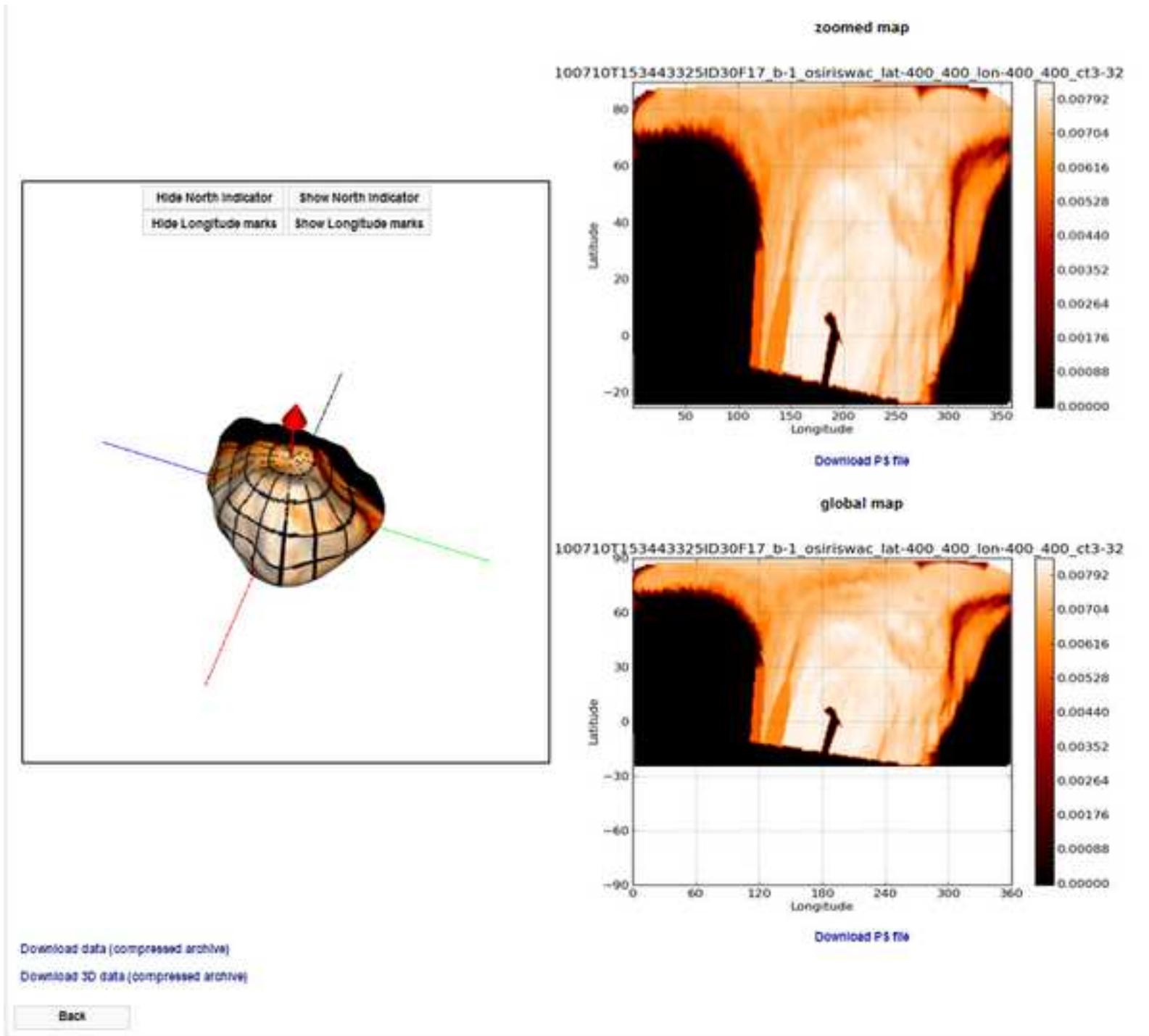

**Figure 6**
**Click here to download high resolution image**

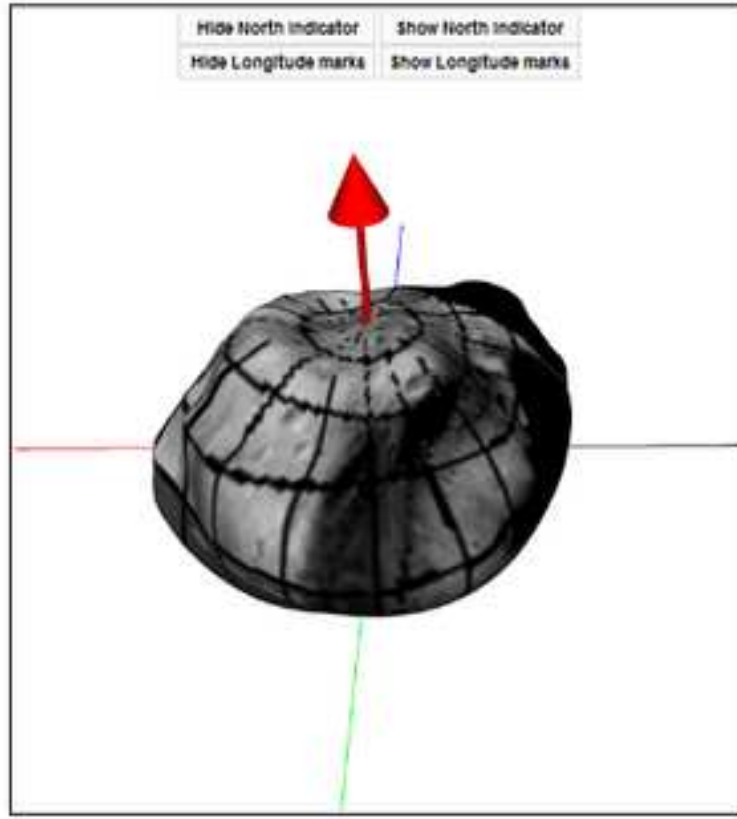
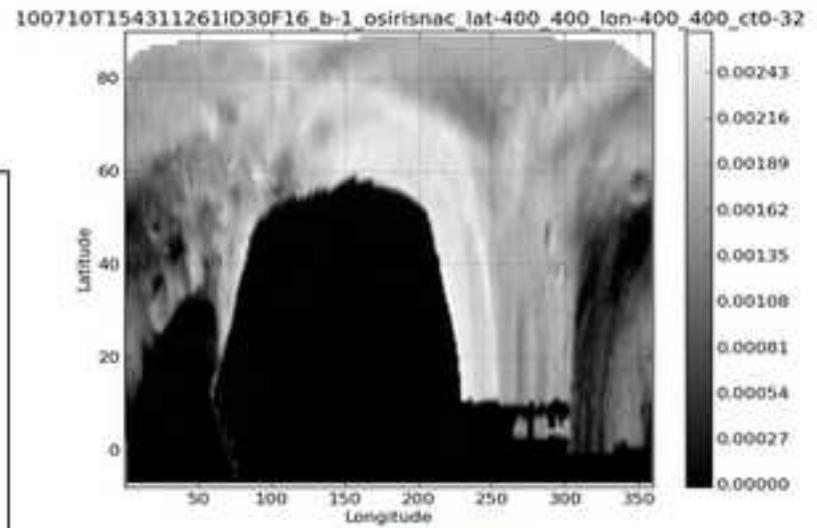
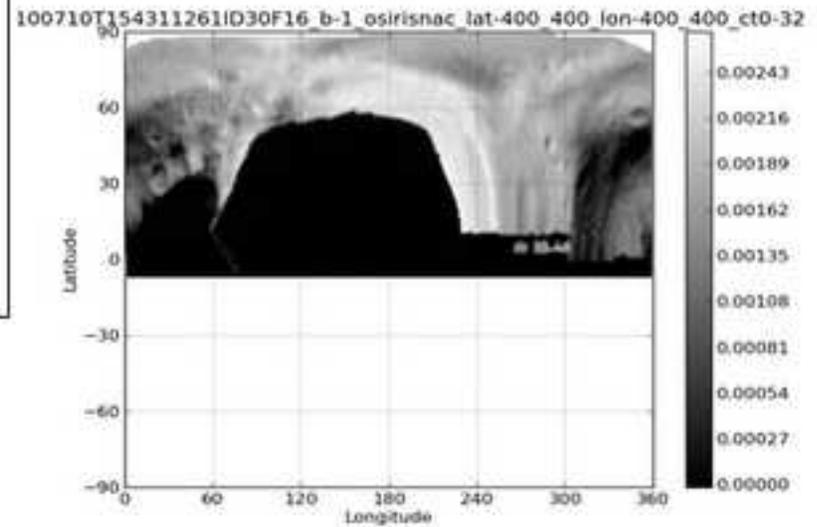

**Figure 7**
**Click here to download high resolution image**

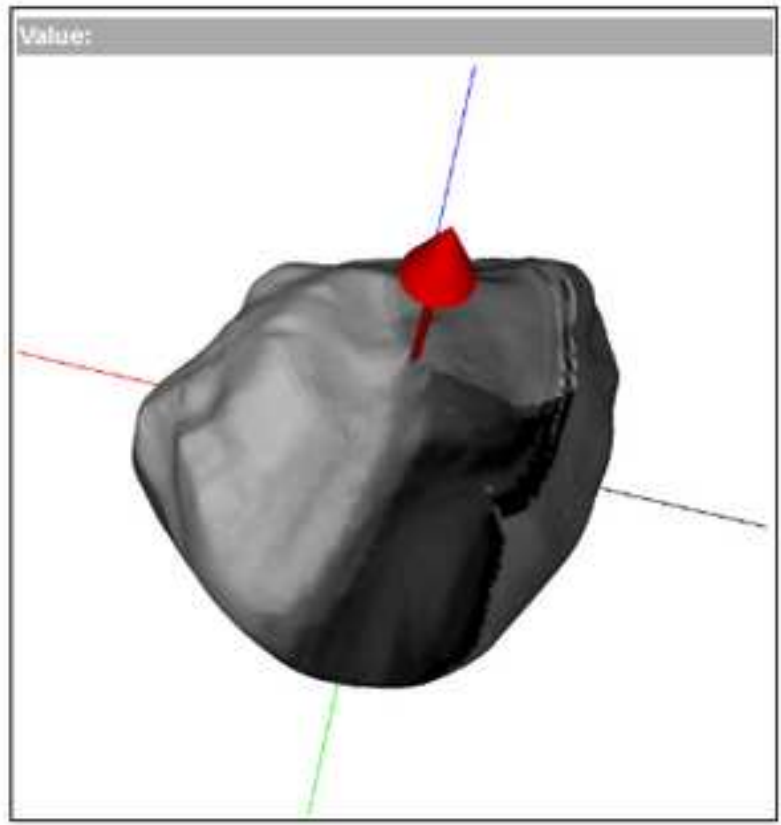
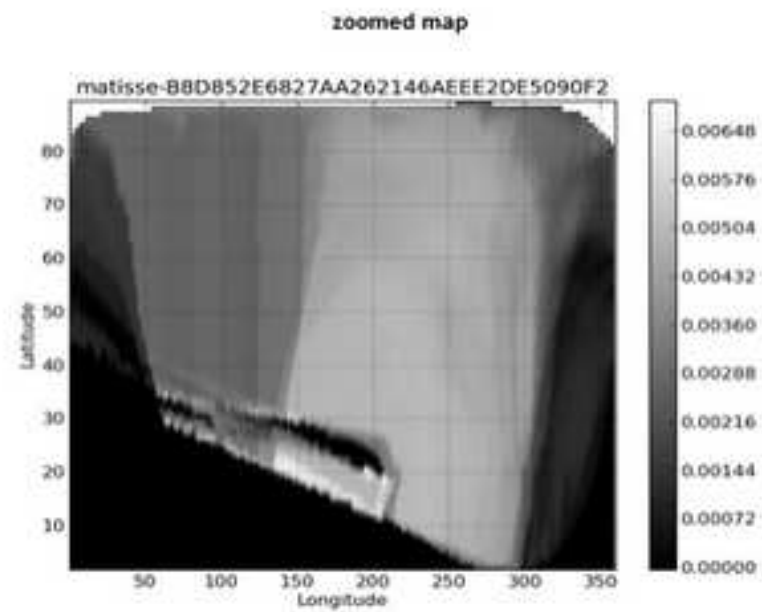
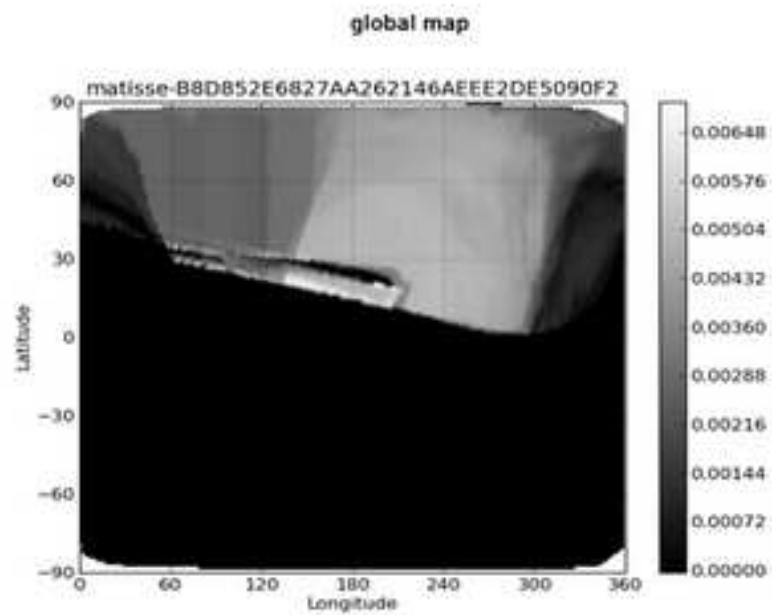

**Figure 8**
**Click here to download high resolution image**

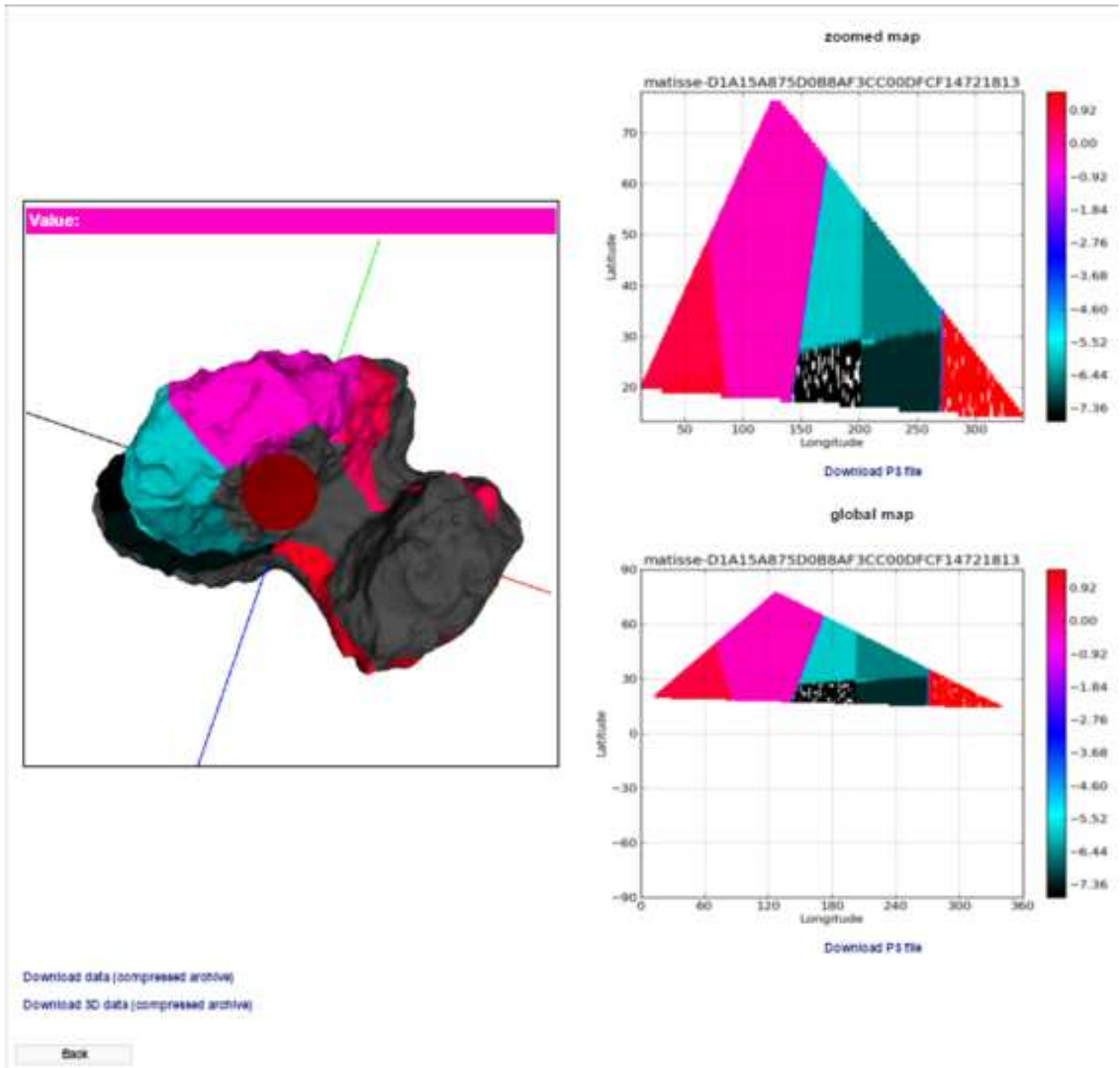

**Figure 9**
**Click here to download high resolution image**

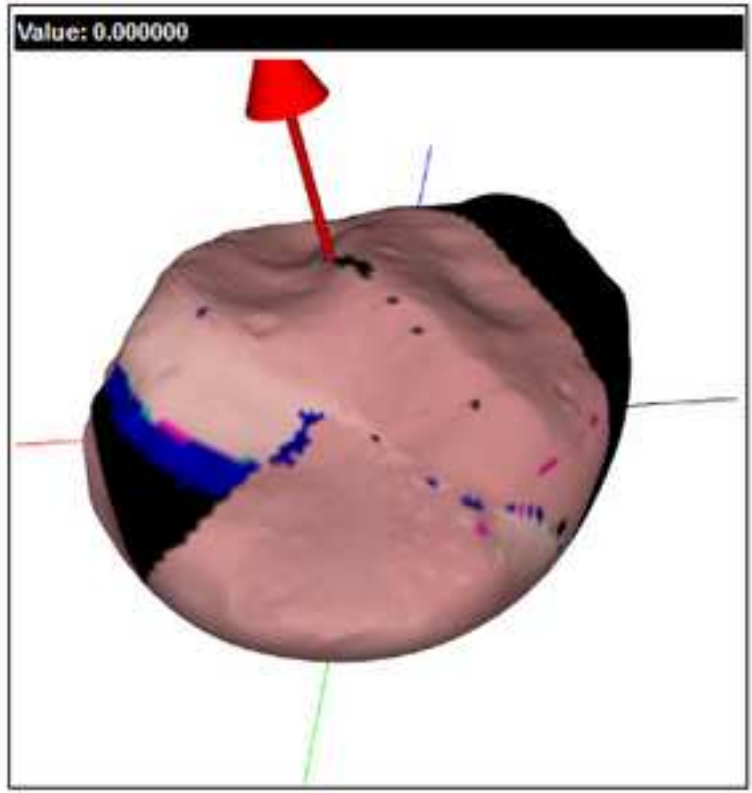
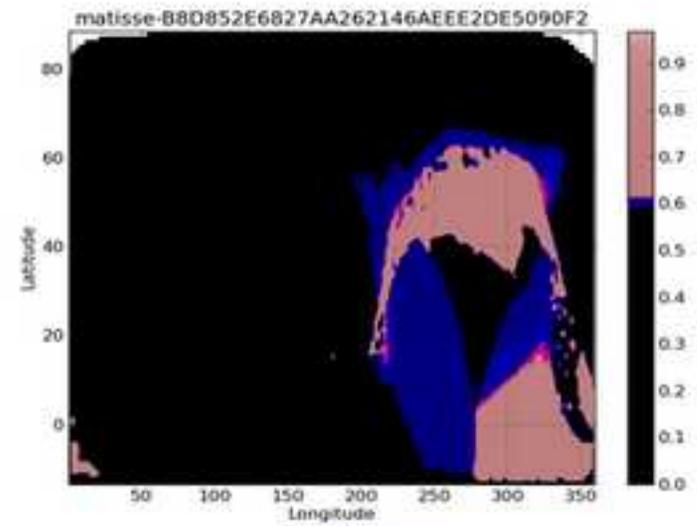
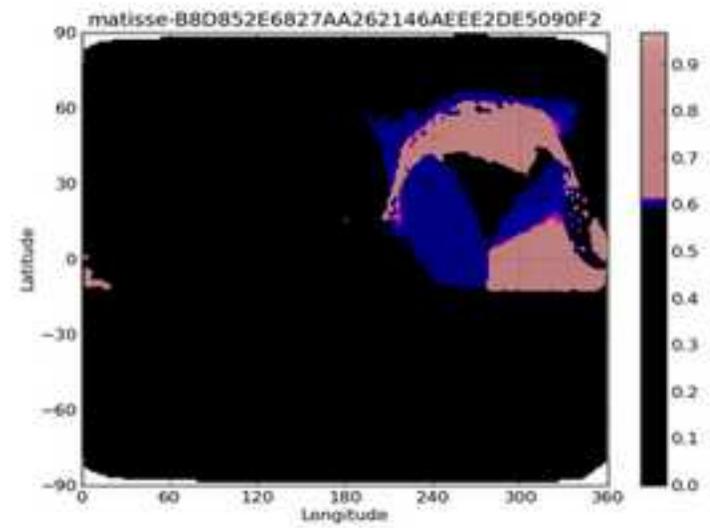

**Figure 10**
**Click here to download high resolution image**

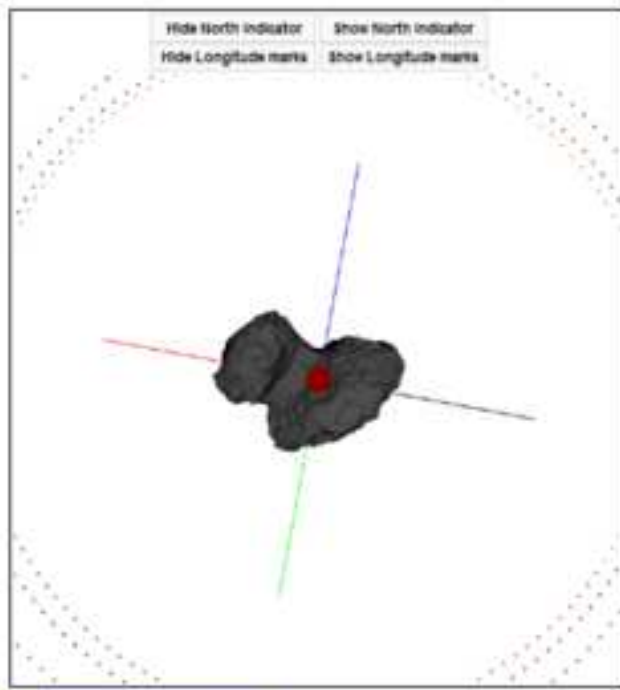 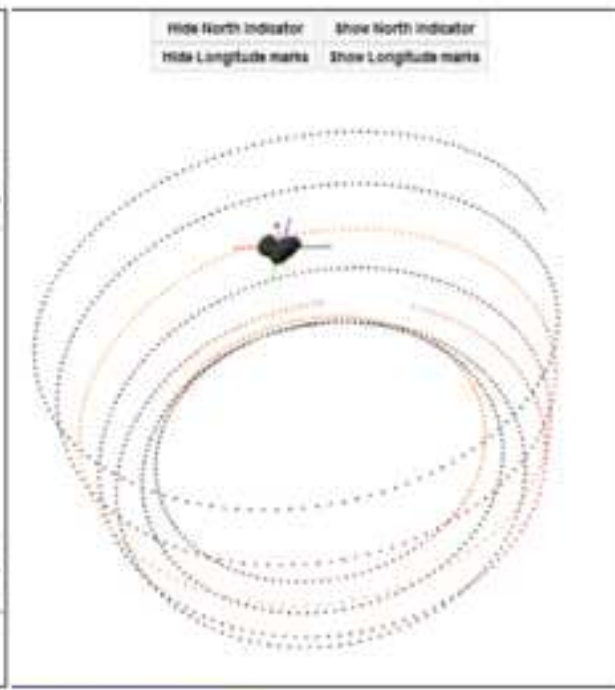 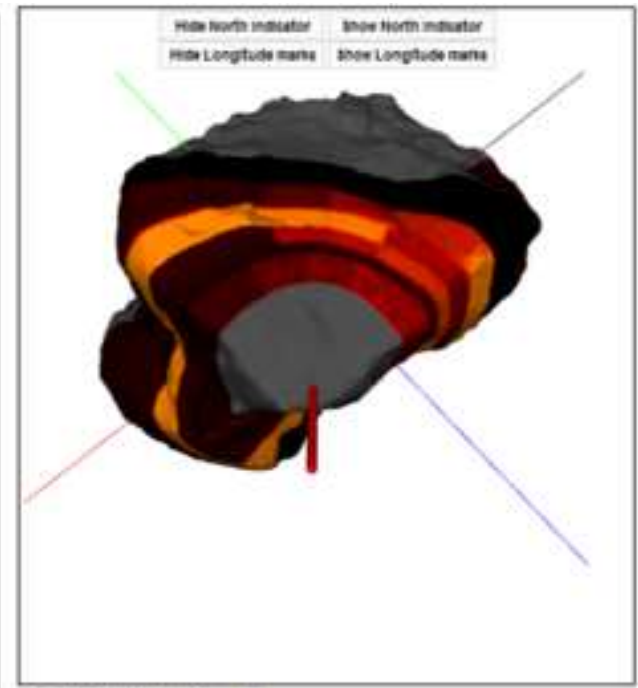

**Figure 11**
**Click here to download high resolution image**

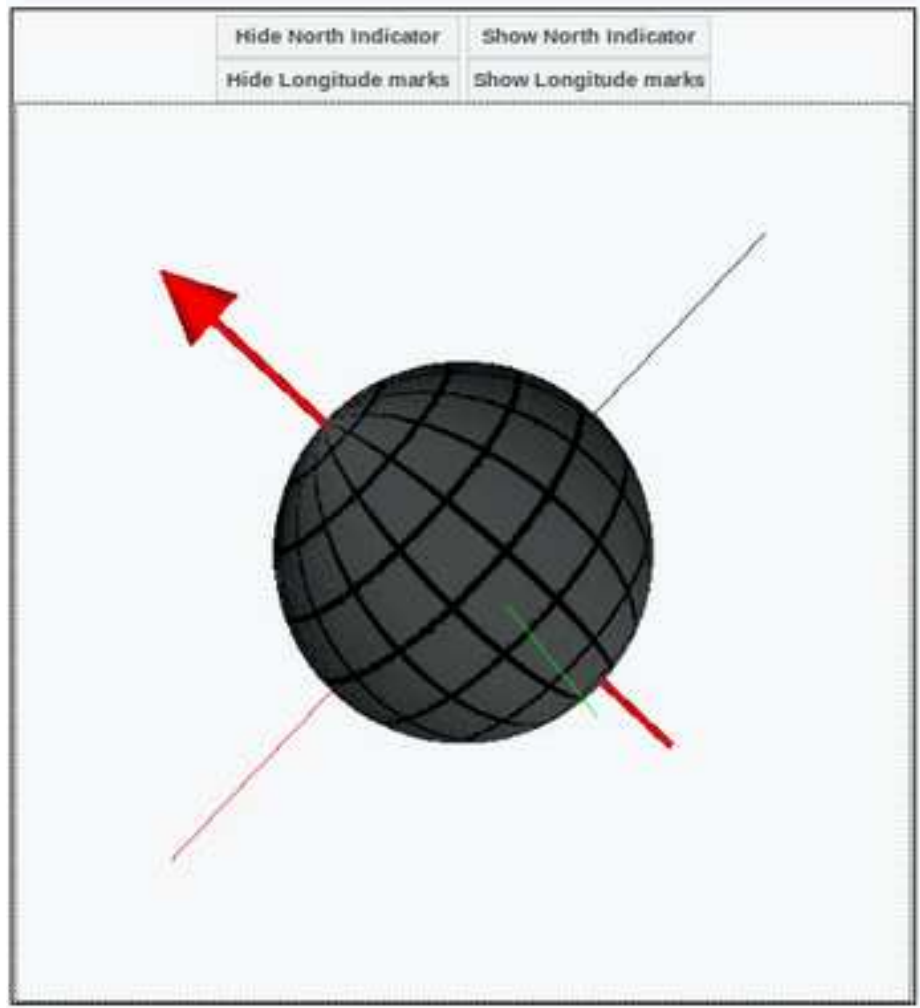
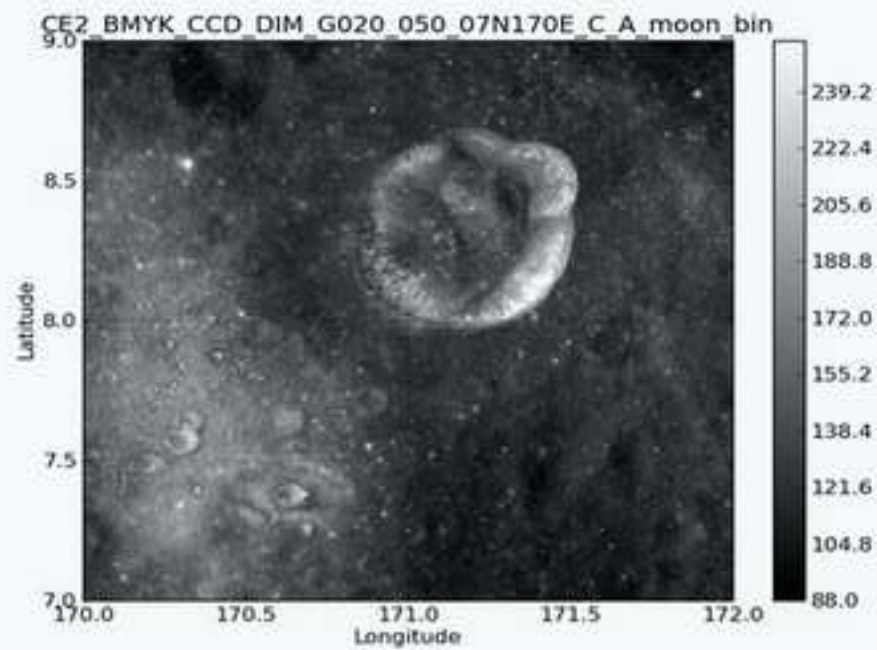

**Figure 12**
Click here to download high resolution image

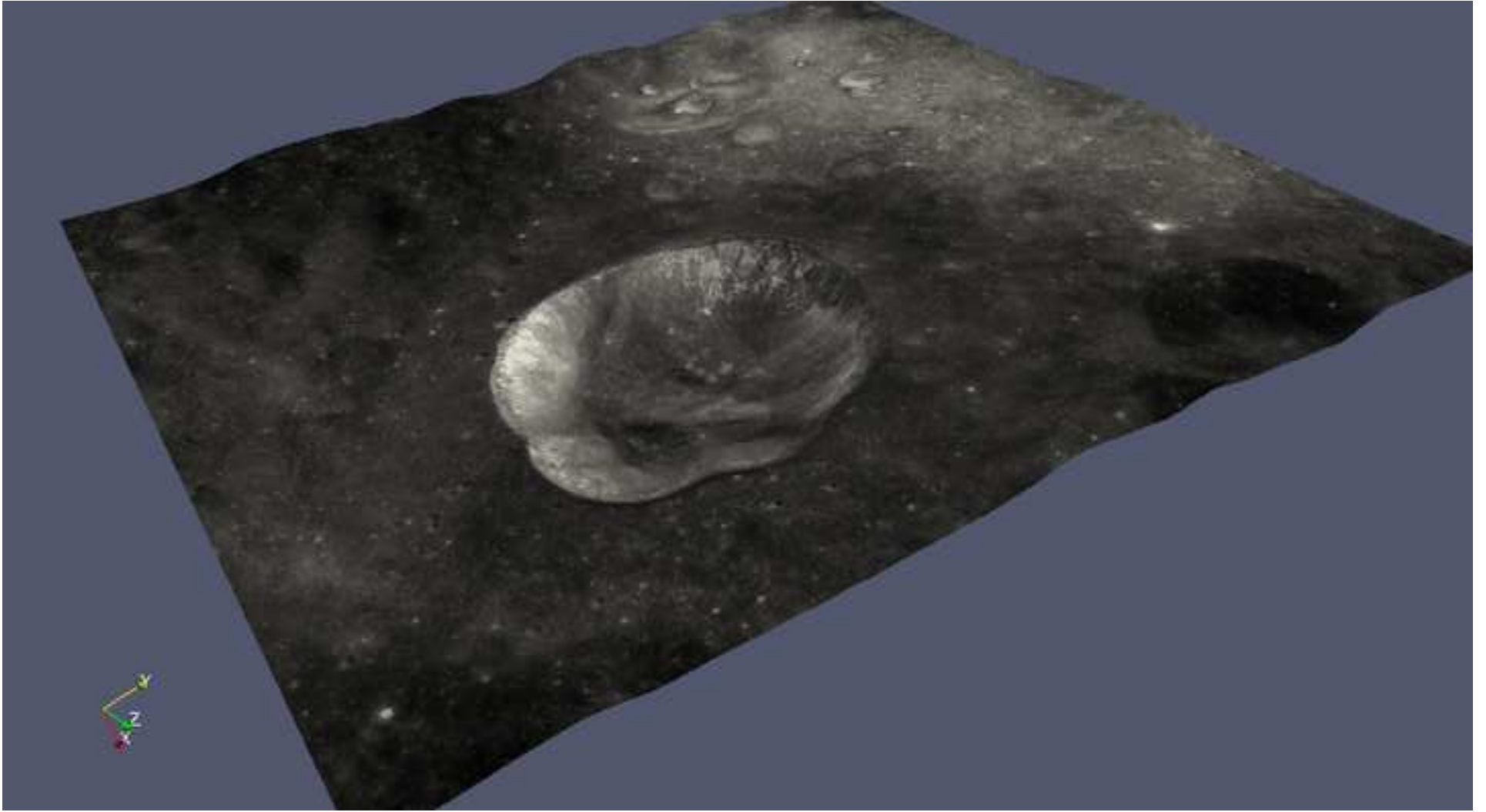

**Tables**

| #  | SOM | MOM ("Submit") | MOM ("Next") |
|----|-----|----------------|--------------|
| 1  | instrScript | instrScript | Operation |
| 2  | MATISSE path | MATISSE path | Palette |
| 3  | Data path | Data path | Number of colors |
| 4  | Filename | Filename | Workdir path |
| 5  | Target | Target | Session ID |
| 6  | Instrument | Instrument | Target |
| 7  | Latitude min | Latitude min | |
| 8  | Longitude min | Longitude min | |
| 9  | Latitude max | Latitude max | |
| 10 | Longitude max | Longitude max | |
| 11 | Wavelength/Channel | Wavelength/Channel | |
| 12 | Palette | Palette | |
| 13 | Number of colors | Number of colors | |
| 14 | Start date | Start date | |
| 15 | End date | End date | |
| 16 | Time step | Time step | |
| 17 | Integration time | Integration time | |
| 18 | Bin size | Bin size | |
| 19 | Incidence angle min | Incidence angle min | |
| 20 | Incidence angle max | Incidence angle max | |
| 21 | Emergence angle min | Emergence angle min | |
| 22 | Emergence angle max | Emergence angle max | |
| 23 | Phase angle min | Phase angle min | |
| 24 | Phase angle max | Phase angle max | |
| 25 | NODATA | NODATA | |
| 26 | | Workdir path | |

**Table 1: Parameters passed by the Java webpage to the bash shell scripts. Parameters from 14 to 18 in columns SOM and MOM ("Submit") are exclusively for the GIADA instrument.**

| #  | Parameter name              | Description                                                              |
|----|-----------------------------|--------------------------------------------------------------------------|
| 1  | Id                          | Autoincremental observation id                                           |
| 2  | observationTypeId           | Observation identification code                                          |
| 3  | missionInstrumentTargetIdTmp| Code that uniquely identifies the combination of Mission, Instrument and Target |
| 4  | file_name                   | Filename                                                                 |
| 5  | Time_min                    | Time minimum value                                                       |
| 6  | Time_max                    | Time maximum value                                                       |
| 7  | C1min                       | Longitude minimum value                                                  |
| 8  | C1max                       | Longitude maximum value                                                  |
| 9  | C2min                       | Latitude minimum value                                                   |
| 10 | C2max                       | Latitude maximum value                                                   |
| 11 | C3min                       | Minimum spacecraft to target distance                                    |
| 12 | C3max                       | Maximum spacecraft to target distance                                    |
| 13 | Incidence_min               | Incidence angle minimum value                                            |
| 14 | Incidence_max               | Incidence angle maximum value                                            |
| 15 | Emergence_min               | Emergence angle minimum value                                            |
| 16 | Emergence_max               | Emergence angle maximum value                                            |
| 17 | Phase_min                   | Phase angle minimum value                                                |
| 18 | Phase_max                   | Phase angle maximum value                                                |
| 19 | Groupname                   | The user group to whom the observation belongs                           |
| 20 | Datapath                    | The directory where the file is stored                                   |

**Table 2: "ObservationMetaData" table columns.**

| Instrument | Mission | Target(s) | # of obs. | Data policy |
|---|---|---|---|---|
| OSIRIS/WAC | Rosetta | Lutetia | 5 | Public |
| OSIRIS/NAC | Rosetta | Lutetia | 1 | Public |
| VIRTIS-M IR | Rosetta | Lutetia / CG | 737 | Public / Private |
| VIRTIS-M VIS | Rosetta | Lutetia / CG | 738 | Public / Private |
| Giada | Rosetta | CG | 4 | Private |
| VIR | Dawn | Vesta | 1 | Private |
| Temp. Simulations | Rosetta | CG | 1153 | Private |
| VIRTIS Retrieved Temp | Rosetta | CG | 585 | Private |
| VIRTIS Retrieved Temp StDev | Rosetta | CG | 585 | Private |
| VIRTIS-M IR Artifact Removed Radiance | Rosetta | CG | 1460 | Private |
| VIRTIS-M VIS Artifact Removed Radiance | Rosetta | CG | 1455 | Private |
| VIRTIS-M IR Reflectance | Rosetta | CG | 1460 | Private |
| VIRTIS-M VIS Reflectance | Rosetta | CG | 1455 | Private |
| Elemental Abundance Map | Chang'e-1 | Moon | 3 | Private |
| DEM | Chang'e-1 | Moon | 188 | Private |
| Ortophoto | Chang'e-1 | Moon | 188 | Private |
| Ortophoto | Chang'e-2 | Moon | 188 | Private |

**Table 3: Composition of the MATISSE database (updated on 21$^{st}$ December 2015). Target names in this table does not exactly correspond to IAU standards only to shorten them in the displayed columns.**

| Section | Target | Instruments | Inc. Angle | Channels | Operation | Part. Size | T. Step | Int. Time |
|---|---|---|---|---|---|---|---|---|
| 4.1 | Lutetia | OSIRIS/NAC | -- | -- | -- | -- | -- | -- |
| 4.2 | Lutetia | OSIRIS/WAC | 0° – 45° | -- | Average | -- | -- | -- |
| 4.3 | 67P | SIMVIRTISD10 | -- | Temp $1^{st}$, Temp $2^{nd}$ layer | Difference | -- | -- | -- |
| 4.4 | Lutetia | VIRTIS-M VIS | -- | 700, 546, 455 nm | RGB | -- | -- | -- |
| 4.5 | 67P | GIADA-Flux | -- | -- | -- | 2.0e16 – 1.0 e16 m | 300 s | 3600 s |
| 4.6 | Moon | CE2 – DOM | -- | -- | -- | -- | -- | -- |

**Table 4: Parameters for the different use cases here shown.**